\documentclass[%
twocolumn,
superscriptaddress,
showpacs,
amsmath,amssymb,amsfonts,
aps,
pra,
floatfix,notitlepage,latexsym
]{revtex4-1}

\usepackage[dvipdfmx]{graphicx}
\usepackage{dcolumn}
\usepackage{bm}
\usepackage{mediabb}
\usepackage[dvipdfmx]{attachfile2}

\begin{document}


\title{Fractional quantum Hall states of dipolar fermions in a strained optical lattice}
\author{Hiroyuki Fujita}
\affiliation{Institute for Solid State Physics, University of Tokyo, 5-1-5 Kashiwanoha, Kashiwa, Chiba 277-8581, Japan}
\author{Yuya O. Nakagawa}
\affiliation{Institute for Solid State Physics, University of Tokyo, 5-1-5 Kashiwanoha, Kashiwa, Chiba 277-8581, Japan}
\author{Yuto Ashida}
\affiliation{Department of Physics, University of Tokyo, 7-3-1 Hongo, Bunkyo-ku, Tokyo 113-0033, Japan}
\author{Shunsuke Furukawa}
\affiliation{Department of Physics, University of Tokyo, 7-3-1 Hongo, Bunkyo-ku, Tokyo 113-0033, Japan}

\date{\today}


\begin{abstract}
We study strongly correlated ground states of dipolar fermions in a honeycomb optical lattice with spatial variations in hopping amplitudes. 
Similar to a strained graphene, such nonuniform hopping amplitudes produce 
valley-dependent pseudomagnetic fields for fermions near the two Dirac points, resulting in the formation of Landau levels. 
The dipole moments polarized perpendicular to the honeycomb plane yield a long-range repulsive interaction. 
By exact diagonalization in the zeroth-Landau-level basis, we show that this repulsive interaction 
stabilizes a variety of valley-polarized fractional quantum Hall states such as Laughlin and composite-fermion states. 
The present system thus offers an intriguing platform for emulating fractional quantum Hall physics in a static optical lattice. 
We calculate the energy gaps above these incompressible states, and discuss the temperature scales required for their experimental realization. 
\end{abstract}

\maketitle

\section{Introduction}

Fractional quantum Hall (FQH) effect \cite{Prange_Girvin_QHE, Yoshioka_QHE}, 
which was first discovered in GaAs/AlGaAs heterostructures \cite{PhysRevLett.48.1559}, 
is a remarkable manifestation of strong correlations between electrons. 
It arises from fractional filling of a massively degenerate Landau level in a high magnetic field, 
where the interaction effect is significantly enhanced. 
Consequently, the ground states are highly entangled in both real and momentum spaces, 
as exemplified by Laughlin wave functions \cite{PhysRevLett.50.1395}. 
FQH states are examples of topologically ordered states of matter with long-range entanglement, 
and exhibit anyonic excitations with fractional charge and statistics \cite{Wen_QFT_MBS}. 
The statistics obeyed by anyons form a representation of the braid group, and can be non-Abelian. 
Possible non-Abelian anyons in the half-filled second Landau level \cite{PhysRevLett.59.1776, MOORE1991362, Stern2008204} 
offer candidate building blocks for a fault-tolerant topological quantum computation \cite{RevModPhys.80.1083}. 


Since the realization of FQH states requires an extremely clean two-dimensional system with high mobility, 
their studies have been limited to silicon, III-V, and oxide heterostructures \cite{PhysRevLett.48.1559, PhysRevLett.59.1776, Nelson1992, DePoortere2002, Tsukazaki2010}  
and graphene \cite{Du2009, Bolotin2009}.
Laser-cooled atomic systems, which have unprecedented cleanness, can offer a new platform for the studies of FQH states \cite{Goldman2016, Bloch2012}. 
While a usual magnetic field does not produce a Lorentz force for neutral atoms, 
different methods of engineering {\it synthetic} magnetic fields that do produce such a force have been developed \cite{RevModPhys.83.1523, Goldman2014_review}. 
Such methods include rotation \cite{PhysRevLett.84.806, Abo-Shaeer476, PhysRevLett.92.040404, cooper_review} 
and optical dressing \cite{PhysRevLett.93.033602, PhysRevA.73.025602, PhysRevA.79.063613, PhysRevA.79.011604, Lin:2009aa} of atoms in continuum 
and laser-induced tunneling in optical lattices 
\cite{PhysRevLett.111.185301, PhysRevLett.111.185302, Aidelsburger:2015aa, Kennedy:2015aa} 
and synthetic dimensions \cite{PhysRevLett.112.043001, Mancini1510, Stuhl1514}. 
On the theoretical side, a variety of FQH states have been predicted to appear in scalar Bose gases in synthetic magnetic fields, 
which include a bosonic Laughlin state \cite{PhysRevLett.80.2265} and non-Abelian Read-Rezayi states \cite{cooper_review, PhysRevB.59.8084, PhysRevLett.87.120405}. 
High controllability of ultracold atoms offers a potential advantage in the manipulation of non-Abelian excitations over solid-state devices. 

While high synthetic magnetic fields have already been realized with the technique of laser-induced tunneling 
\cite{PhysRevLett.111.185301, PhysRevLett.111.185302, Aidelsburger:2015aa, Kennedy:2015aa, Mancini1510, Stuhl1514}, 
the Raman processes used in this technique involve heating of the system, 
which crucially limits the time scale of experiments. 
Recently, Tian, Endres, and Pekker \cite{PhysRevLett.115.236803} have proposed an interesting scheme that is free from this difficulty. 
Their theoretical proposal is inspired by the fact that 
in graphene \cite{Guinea:2010aa, Levy544, doi:10.1021/nl1018063} and molecular graphene \cite{Gomes2012}, 
nonuniform strain induces valley-dependent high pseudomagnetic fields for fermions near the two Dirac points \cite{RevModPhys.81.109}.  
The authors of Ref.~\cite{PhysRevLett.115.236803} have proposed a method of generating spatially varying hopping amplitudes in a honeycomb optical lattice, 
which can mimic these systems. 
It is based on a simple configuration where three Gaussian laser beams intersect at 120$^\circ$ 
but their centers are displaced from the center of the system. 
This scheme can realize quasiuniform high pseudomagnetic fields in a static optical lattice, 
and significantly enlarge the time scale of experiments. 

It is interesting to ask what quantum phases emerge by loading interacting fermions in such a ``strained'' honeycomb optical lattice. 
For strained graphene, where electrons interact via a Coulomb interaction, 
the emergence of valley-polarized (fractional) quantum Hall states and valley-symmetric topological states has been discussed 
\cite{PhysRevLett.108.266801, PhysRevLett.109.066802}. 
For ultracold spin-$\frac12$ fermionic atoms, the dominance of an intercomponent $s$-wave interaction is likely to lead to the spontaneous spin polarization; 
the resulting system is essentially noninteracting due to the absence of an intracomponent $s$-wave interaction, and cannot stabilize a topologically ordered state. 
By contrast, if the fermions possess large electric or magnetic dipole moments \cite{Baranov2012_review}, 
they interact via a long-range interaction even when the spin state is polarized. 
There has recently been a remarkable progress in the creation and manipulation of dipolar Fermi gases. 
Fermionic polar molecules such as $^{40}$K$^{87}$Rb \cite{B911779B, Ni231, B821298H, PhysRevLett.104.030402} 
and ${}^{23}$Na${}^{40}$K \cite{PhysRevLett.114.205302, ParkWillZwierlein2014} have been prepared in their absolute ground states 
while magnetic atoms such as $^{161}$Dy \cite{PhysRevLett.108.215301}, $^{167}$Er \cite{PhysRevLett.112.010404}, 
and $^{53}$Cr \cite{PhysRevA.91.011603} have been brought to Fermi degeneracy. 



In this paper, we study strongly correlated ground states of dipolar fermions in a strained honeycomb optical lattice, 
which can be realized with the scheme of Ref.~\cite{PhysRevLett.115.236803}. 
The dipole moments are taken to be polarized perpendicular to the honeycomb plane, yielding a long-range repulsive interaction. 
The low-energy effective theory of this system is given by interacting Dirac fermions near two valleys in mutually antiparallel magnetic fields. 
We simulate this theory by exact diagonalization (ED) in the zeroth-Landau-level (ZLL) basis in a spherical geometry. 
We find that there appear a variety of valley-polarized FQH states 
such as Laughlin \cite{PhysRevLett.50.1395} and composite-fermion states \cite{Jain_CompositeFermion, PhysRevLett.63.199, PhysRevB.41.7653} of particles and holes. 
The present system thus offers an intriguing platform for emulating FQH physics in a static optical lattice. 
We calculate the energy gaps above these incompressible states, and discuss the temperature scales required for their experimental realization. 

The rest of the paper is organized as follows.  
In Sec.\ \ref{sec. 2}, we describe our system and explain how spatially varying hopping amplitudes in a honeycomb optical lattice generate pseudomagnetic fields. 
We then derive a low-energy effective theory of this system, which has the form of interacting two-species Dirac fermions in antiparallel magnetic fields. 
In Sec.\ \ref{sec. 3}, we formulate the problem using the ZLL basis in the spherical geometry, which is useful in numerical analyses. 
In Sec.\ \ref{sec. 4}, we present our ED results. 
In particular, we perform an extensive search for FQH states, and calculate the energy gaps above these ground states. 
We discuss the possibility of realizing these states in a particular optical lattice setup. 
In Sec.\ \ref{sec. 5}, we present a summary of this paper, and discuss an outlook for future studies. 
In Appendix \ref{sec:calc_pseudopotential}, we give details of the calculation of pseudopotentials. 



\section{Dipolar fermions in a strained honeycomb optical lattice}\label{sec. 2}

\newcommand{\Hcal}{{\cal H}}
\newcommand{\Acal}{{\cal A}}
\newcommand{\kin}{\mathrm{kin}}
\newcommand{\inter}{\mathrm{int}}
\newcommand{\Hcalt}{\tilde{\cal H}}
\newcommand{\Ncal}{{\cal N}}
\newcommand{\Ncalt}{\tilde{\cal N}}
\newcommand{\ub}{\bar{u}}
\newcommand{\vb}{\bar{v}}
\newcommand{\St}{\tilde{S}}
\newcommand{\nut}{\widetilde{\nu}}

\newcommand{\deltav}{\bm{\delta}}
\newcommand{\tauv}{\bm{\tau}}
\newcommand{\sigmav}{\bm{\sigma}}
\newcommand{\Lambdav}{{\bm{\Lambda}}}
\newcommand{\piv}{\bm{\pi}}

\newcommand{\ev}{{\bm{e}}}
\newcommand{\rv}{{\bm{r}}}
\newcommand{\kv}{{\bm{k}}}
\newcommand{\Kv}{{\bm{K}}}
\newcommand{\qv}{{\bm{q}}}
\newcommand{\pv}{{\bm{p}}}
\newcommand{\Av}{{\bm{A}}}
\newcommand{\Lv}{{\bm{L}}}
\newcommand{\Jv}{{\bm{J}}}

\newcommand{\bra}[1]{\langle #1|}
\newcommand{\ket}[1]{| #1 \rangle}
\newcommand{\bracket}[2]{\langle #1|#2 \rangle}
\newcommand{\bralr}[1]{\left\langle #1 \right|}
\newcommand{\ketlr}[1]{\left| #1 \right\rangle}
\newcommand{\bbra}[1]{\langle\!\langle #1|}
\newcommand{\kett}[1]{|#1 \rangle\!\rangle}
\newcommand{\bbrackett}[2]{\langle\!\langle #1|#2 \rangle\!\rangle}
\newcommand{\ua}{\uparrow}
\newcommand{\da}{\downarrow}

We consider a system of fermions loaded into a honeycomb optical lattice 
with an effective ``strain'' due to spatially varying hopping amplitudes. 
Each fermionic atom or molecule possesses a electric or magnetic dipole moment 
polarized perpendicular to the honeycomb plane, 
yielding a long-range dipole-dipole interaction $V(r)=Cr^{-3}$, where $C$ is a constant. 
We review how spatially varying hopping amplitudes produce 
valley-dependent pseudomagnetic fields for fermions near the two Dirac points. 
We then introduce the continuum description of the system 
in terms of interacting two-species Dirac fermions in antiparallel magnetic fields. 

The honeycomb lattice consists of two sublattices $A$ and $B$. 
We introduce three vectors
\begin{equation}\label{eq:deltav}
 \deltav_1=\frac{a}{2} (\sqrt{3}\ev_x+\ev_y),~
 \deltav_2=\frac{a}{2} (-\sqrt{3}\ev_x+\ev_y),~
 \deltav_3=-a\ev_y,
\end{equation}
which connect any $A$ site to its three neighboring $B$ sites. 
Here, $a$ is the length of a nearest-neighbor bond, 
and $\ev_x$ and $\ev_y$ are the unit vectors along $x$ and $y$ directions, respectively. 
The triangular Bravais lattice is generated by the basis vectors $\bm{a}_1=\deltav_1-\deltav_2$ and $\bm{a}_2=\deltav_1-\deltav_3$, 
and the area of the unit cell is ${\cal A}_c=|\bm{a}_1\times\bm{a}_2|=3\sqrt{3}a^2/2$. 

For a sufficiently deep optical lattice, the kinetic part $H_\kin$ of the Hamiltonian is well described by a tight-binding model in a honeycomb lattice. 
We first consider a spatially uniform optical lattice with hopping amplitudes $t_j(>0)$ along $\deltav_j~(j=1,2,3)$ 
\footnote{Hopping amplitudes usually appear with a negative sign in optical lattices, 
but in the honeycomb case, it can be removed by the gauge transformation for the fermionic operators on the $B$ sublattices: 
$c(\rv)\to-c(\rv)~(\rv\in B)$.}. 
In this case, $H_\kin$ is given by
\begin{align}
&H_\kin = \sum_\kv \left( c_A^\dagger (\kv), c_B^\dagger (\kv) \right) 
\begin{pmatrix}  0 &   f^*(\kv) \\ f(\kv) & 0 \end{pmatrix}
 \begin{pmatrix} c_A(\kv) \\ c_B(\kv) \end{pmatrix},\\
&f(\kv)=\sum_j t_j e^{-i\kv\cdot\deltav_j},
\end{align}
where $c_X(\bm{k})$ annihilates a fermion with wave vector $\bm{k}$ on the sublattice $X(=A,B)$. 
When $t_1=t_2=t_3$, the energy bands $\pm |f(\kv)|$ exhibit two Dirac cones 
at the two Brillouin zone corners $\Kv_\xi=\xi\Kv=-\xi (4\pi/3\sqrt{3}a)\ev_x$, where $\xi=\pm$ is the valley index. 
When $t_j$'s are not equal, the Dirac points are shifted from $\Kv_\xi$. 
To see it, we set $\kv=\xi\Kv+\qv$ and  expand $f(\kv)$ in terms of $\qv a$. 
Assuming $t_j\approx t~(j=1,2,3)$, we find
\begin{equation}\label{eq:f_shift}
 f(\xi\Kv+\qv)\approx v_F \left[ \xi (\hbar q_x-\xi A_x) +  i(\hbar q_y-\xi A_y) \right],
\end{equation}
where $v_F=3ta/2\hbar$ is the velocity of the Dirac fermions, and
\begin{equation}
 v_F A_x=\frac12(t_1+t_2)-t_3, ~v_FA_y=\frac{\sqrt{3}}{2} (t_2-t_1). 
\end{equation}
As seen in Eq.~\eqref{eq:f_shift}, the two Dirac points shift in mutually opposite directions by the vectors $\xi\Av/\hbar$. 
When the Fermi level is close to the Dirac points, 
$H_\kin$ can be effectively described at low energies by two-species massless Dirac fermions as
\begin{align}
H_\kin
&=\sum_{\xi=\pm} \sum_{\qv:|\qv|<\Delta} \left( c_A^\dagger(\xi\Kv+\qv),c_B^\dagger(\xi\Kv+\qv) \right) \notag\\
&~~~~~~~~\times v_F(\hbar\qv-\xi\Av)\cdot\tauv_\xi 
 \begin{pmatrix} c_A(\xi\Kv+\qv) \\ c_B(\xi\Kv+\qv) \end{pmatrix} \label{eq:H_Dirac_discrete}\\
&\to v_F \sum_{\xi=\pm} \int d^2\rv \Psi_\xi^\dagger (\rv) (\pv-\xi\Av) \cdot \tauv_\xi \Psi_\xi (\rv), \label{eq:H_Dirac}
\end{align}
where $\Delta$ is a high-wave-number cutoff, $\pv=-i\hbar(\partial_x,\partial_y)$ is the momentum operator, 
and $\tauv_\xi=(\xi \sigma^x,\sigma^y)$ with $(\sigma^x,\sigma^y,\sigma^z)$ being the Pauli matrices. 
When going from Eq.~\eqref{eq:H_Dirac_discrete} to Eq.~\eqref{eq:H_Dirac}, we have performed replacements
\begin{equation}
 \sum_{\qv:|\qv|<\Delta} \to \int\frac{{\cal A}_c d^2\qv}{(2\pi)^2},~
 \begin{pmatrix} c_A(\xi\Kv+\qv) \\ c_B(\xi\Kv+\qv) \end{pmatrix} \to \frac{\Psi_\xi(\qv)}{\sqrt{{\cal A}_c}},
\end{equation}
and the Fourier transformation
\begin{equation}
 \Psi_\xi(\rv)= \int \frac{d^2\qv}{(2\pi)^2} \Psi_\xi (\qv) e^{i\qv\cdot\rv}.
\end{equation}
The fermionic operator $c(\rv)$ on the original lattice site $\rv\in X(=A,B)$ 
is related to $\Psi_\xi(\rv)={}^t(\Psi_{\xi,A}(\rv),\Psi_{\xi,B}(\rv))$ as
\begin{equation}\label{eq:c_Psi}
\begin{split}
c(\rv) 
&= \sum_{\xi=\pm} \sum_{\qv:|\qv|<\Delta} c_X (\xi\Kv+\qv) e^{i(\xi\Kv+\qv)\cdot\rv}\\
&\to \sqrt{\Acal_c} \sum_{\xi=\pm} e^{i\xi\Kv\cdot\rv} \Psi_{\xi,X}(\rv) .
\end{split}
\end{equation}

When the hopping amplitudes vary slowly in space, 
the shift $\xi \Av$ in the Dirac Hamiltonian \eqref{eq:H_Dirac} also varies spatially and plays a role of a pseudovector potential. 
Tian {\it et al.} \cite{PhysRevLett.115.236803} have proposed that such spatially varying hopping amplitudes $t_j(\rv)~(j=1,2,3)$ 
can be generated in a honeycomb optical lattice 
by starting from a standard configuration of three Gaussian laser beams intersecting at 120$^\circ$ \cite{Soltan-Panahi2011, Duca288}
and displacing the centers of the beams from the center of the systems. 
The induced pseudovector potentials $\xi\Av(\rv)$ are shown to lead to quasiuniform pseudomagnetic fields $\xi B(\rv)=\xi [\partial_x A_y(\rv)-\partial_y A_x(\rv)]$ 
for fermions near the two valleys. 
The mutually opposite signs of the pseudomagnetic fields for the two valleys 
come from the fact that the nonuniform hopping amplitudes do not break the time-reversal symmetry. 
We note that here time reversal is defined as the complex conjugation operator for polarized fermions (and does not involve a spin rotation). 

Next we consider the interaction part of the Hamiltonian, 
\begin{equation}
 H_\inter = \frac12 \sum_{\rv,\rv'} V(|\rv-\rv'|) :n(\rv)n(\rv'):,
\end{equation}
where $n(\rv)=c^\dagger(\rv)c(\rv)$ is the number operator at the site $\rv$, and the colons $:\cdot:$ indicate the normal ordering. 
By performing replacement $\sum_\rv=\sum_{X=A,B}\sum_{\rv\in X}\to \sum_{X} \int d^2\rv/\Acal_c$ and using Eq.~\eqref{eq:c_Psi}, 
we obtain the effective interactions between Dirac fermions as
\begin{equation}
\begin{split}
H_\inter
\to &\sum_{\xi_1,\xi_2,\eta_1,\eta_2=\pm} \int d^2\rv d^2\rv' ~ V(|\rv-\rv'|) \\
&~~~~~~\times e^{-i(\xi_1-\xi_2)\Kv\cdot\rv-i(\eta_1-\eta_2)\Kv\cdot\rv'}\\
&~~~~~~\times :\Psi_{\xi_1}^\dagger (\rv) \Psi_{\xi_2}(\rv) \Psi_{\eta_1}^\dagger (\rv') \Psi_{\eta_2}(\rv'): .
\end{split}
\end{equation}
Here, valley-converting processes with $\xi_1\ne\xi_2$ or $\eta_1\ne\eta_2$ 
involve highly oscillating factors $e^{\pm 2i\Kv\cdot\rv}$ or $e^{\pm 2i\Kv\cdot\rv'}$, 
and can be neglected if the interaction $V(\rv-\rv')$ varies slowly over the scale of the lattice constant. 
In such a case, the interaction Hamiltonian can be recast 
into the sum of intra-valley and inter-valley density-density interactions as
\begin{equation}
\begin{split}
 &H_\inter = \sum_{\xi,\eta=\pm} H_\inter^{(\xi,\eta)}, \label{eq:H_inter_eff}\\
 &H_\inter^{(\xi,\eta)}=\frac12 \int d^2\rv d^2\rv' ~V(|\rv-\rv'|) :\rho_\xi (\rv) \rho_\eta (\rv') :,
\end{split}
\end{equation}
where $\rho_\xi(\rv)=\Psi_\xi^\dagger (\rv)\Psi_\xi (\rv)$ is the density operator for the valley $\xi=\pm$. 

  \begin{figure}
   \centering
    \includegraphics[width=0.5\textwidth]{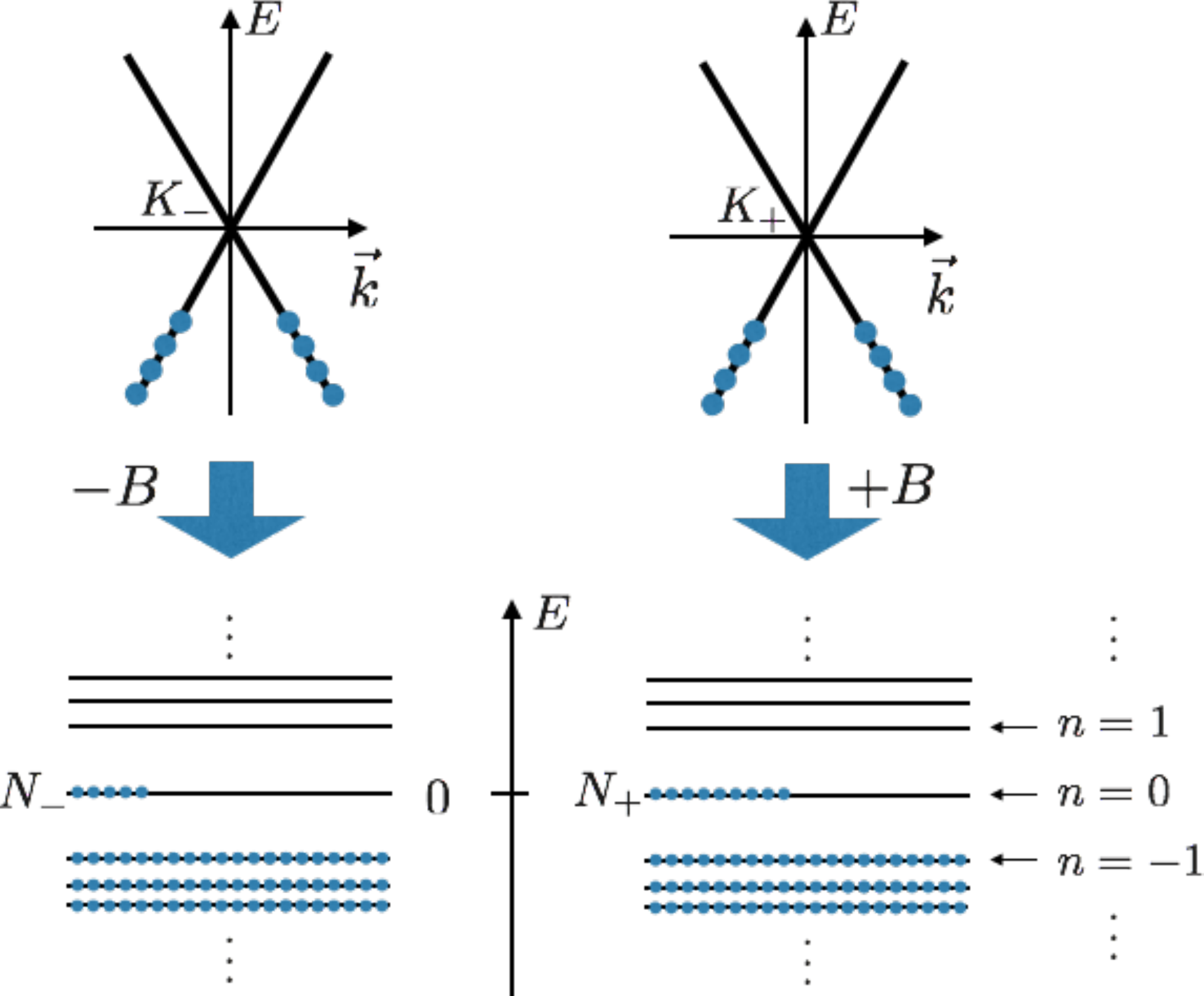}
\caption{(Color online) 
Top: Dirac spectra near the valleys $\Kv_\pm$. 
Bottom: Relativistic Landau levels \eqref{eq:spec_disc} in valley-dependent pseudomagnetic fields $\pm B$. 
Each level at each valley is $N_\phi$-fold degenerate. 
We consider the case when the Fermi level lies near the ZLL $n=0$ and this level is partially populated. 
The number of fermions in the ZLL at the valley $\Kv_\pm$ is denoted by $N_\pm$. 
}
          \label{Landau_levels}
    \end{figure}

At each valley $\Kv_\xi$, a spatially uniform pseudomagnetic field $\xi B$ 
leads to the formation of relativistic Landau levels \cite{Berestetskii_book, Novoselov2005} 
\begin{equation}\label{eq:spec_disc}
 \epsilon_n = \frac{\sqrt{2}\hbar v_F}{l_B} \mathrm{sgn}(n) \sqrt{|n|},~n\in\mathbb{Z}.
\end{equation}
Each level has the degeneracy $N_\phi=\Acal/2\pi l_B^2$, where $\Acal$ is the area of the system. 
Below we consider the case when the Fermi level lies near the ZLL $n=0$ and this level is partially populated as in Fig.~\ref{Landau_levels}. 
When the interaction energy scale is much smaller than the Landau-level spacing $\sqrt{2}\hbar v_F/l_B$, 
we can analyze the interaction Hamiltonian \eqref{eq:H_inter_eff} within the restricted manifold spanned by the ZLL states. 
The number of fermions, $N_\xi$, in the ZLL is independently conserved at each valley $\Kv_\xi$ 
since $H_\inter$ does not involve inter-valley tunneling within our approximation in Eq.~\eqref{eq:H_inter_eff}. 
Exact diagonzalization calculation can thus be performed separately in each sector with fixed $N_\pm$.  
Similar to the case of graphene, we define the filling factor $\nu$ as
\begin{equation}\label{eq:nu_N}
 \nu=-1+\nut=\frac{N_++N_--N_\phi}{N_\phi}, 
\end{equation}
which ranges over $-1<\nu<1$ in the present case. 
The case of $\nu=0$ corresponds to the half-filled ZLL. 
Because the particle-hole transformation relates the physics for $\pm|\nu|$, 
we focus on the case of $-1<\nu<0$ (i.e., $0<\nut<1$).

\section{Spherical geometry}\label{sec. 3}

To analyze the interaction Hamiltonian \eqref{eq:H_inter_eff} in the ZLL, 
it is useful to adopt the spherical geometry \cite{PhysRevLett.51.605,PhysRevB.34.2670}, which is uniform and has no edge. 
In Sec.~\ref{sec:ZLL}, we review the relativistic Landau model on a sphere, which has been solved in Refs.~\cite{Kolesnikov2006,Deguchi01062006}.
We describe the derivation of the single-particle eigenstates in the ZLL, based on an algebraic method formulated recently by Hasebe \cite{Hasebe2015}. 
In Sec.~\ref{sec:pseudopotential}, we construct Haldane's pseudopotentials \cite{PhysRevLett.51.605} in the ZLL for a power-law-decaying interaction, 
with particular focus on the case of a dipole-dipole interaction. 


\subsection{Single-particle eigenstates} \label{sec:ZLL}

We consider two-species Dirac fermions labeled by the valley index $\xi=\pm$ and subject to antiparallel pseudomagnetic fields on a sphere. 
Each species has two pseudospin states, which correspond to the two sublattices of the honeycomb lattice. 
We introduce the polar coordinates $(r,\theta,\phi)$ and the associated unit vectors $\ev_r,\ev_\theta,\ev_\phi$. 
We place magnetic monopoles of valley-dependent integer charges $\xi N_\phi\equiv \xi (2S)$ 
(in units of flux quantum $2\pi\hbar$) at the center of the sphere. 
These monopoles produce uniform magnetic fields $\xi B\ev_r$ on the sphere of radius $R=l_B \sqrt{S}$, 
where $l_B=\sqrt{\hbar/B}$ is the magnetic length. 
The corresponding vector potentials in the Schwinger gauge are given by $\xi \Av(\rv)$, where 
\begin{align}\label{eq:A_monopole}
\Av(\rv) =-2\pi\hbar N_\phi \frac{\cot\theta}{4\pi r} \ev_\phi = -\frac{\hbar S\cot\theta}{r} \ev_\phi.
\end{align}

As an analogue of the Dirac Hamiltonian in Eq.~\eqref{eq:H_Dirac}, 
we consider the following single-particle Hamiltonian on a sphere: 
\begin{equation}\label{eq:Hcal_def}
 \Hcal_\xi = v_F \left( \pi_\xi^\theta \tau_\xi^x + \pi_\xi^\phi \tau_\xi^y \right),
\end{equation}
where 
\begin{equation}\label{eq:pi_def}
 \ev_\theta\pi_\xi^\theta+\ev_\phi \pi_\xi^\phi 
 = \left[ -i\hbar\bm{\nabla} -\xi\Av(\rv) - \ev_\phi \frac{\hbar}{R} \frac{\tau_\xi^z}{2}\cot\theta \right] \bigg|_{r=R},
\end{equation}
is an analogue of the dynamical momentum for the relativistic problem, and $\tauv_\xi=(\xi\sigma^x,\sigma^y,\xi\sigma^z)$.
Here, the last term in Eq.~\eqref{eq:pi_def} originates from the spin connection \cite{Kolesnikov2006,Deguchi01062006,Hasebe2015}, 
and has the effect of modifying the monopole charge $\xi(2S)$ by $\xi(\mp 1)$. 
Using Eq.~\eqref{eq:A_monopole} and the representation of $\bm{\nabla}$ in the spherical coordinate, we obtain
\begin{equation}
 \pi_\xi^\theta=\frac{\hbar}{R} (-i \partial_\theta),~~
 \pi_\xi^\phi = \frac{\hbar}{R} \left[ \frac{-i}{\sin\theta}\partial_\phi + \xi \left( S-\frac{\sigma^z}{2} \right) \cot\theta\right].
\end{equation}

To reveal the algebraic aspect of the problem, 
it is useful to introduce the edth differential operators \cite{NewmanPenrose1966,Dray1985,Dray1986}
\begin{equation}\label{eq:edth_def}
\begin{split}
 \eth_+^{(S)}
 &\equiv (\sin\theta)^{ S} (\partial_\theta+\frac{i}{\sin\theta}\partial_\phi) (\sin\theta)^{-S}\\
 &=\partial_\theta- S\cot\theta+\frac{i}{\sin\theta}\partial_\phi,\\
 \eth_-^{(S)} 
 &\equiv (\sin\theta)^{- S} (\partial_\theta-\frac{i}{\sin\theta}\partial_\phi) (\sin\theta)^{S}\\
 &=\partial_\theta+  S\cot\theta-\frac{i}{\sin\theta}\partial_\phi,
\end{split}
\end{equation}
and the orbital angular momentum operator
\begin{equation}
 \Lv^{(S)} = \rv\times \left( \pv+\frac{\hbar S\cot\theta}{r}\ev_\phi \right) -\hbar S\ev_r.
\end{equation}
Here, $\Lv^{(S)}$ is the generator of the spherical symmetry in the non-relativistic Landau problem on a sphere
with a monopole charge $2S$ \cite{PhysRevLett.51.605,PhysRevB.34.2670}, 
and obeys the standard algebra of an angular momentum. 
Furthermore, the edth and angular momentum operators obey the following algebra:
\begin{subequations}
\begin{align}
 &\eth_+^{(S-1)} \eth_-^{(S)} - \eth_-^{(S+1)} \eth_+^{(S)}=-2S, \label{eq:edth_al_p}\\
 &\eth_+^{(S-1)} \eth_-^{(S)} + \eth_-^{(S+1)} \eth_+^{(S)}=-2({\Lv^{(S)}}^2-S^2), \label{eq:edth_al_m}\\
 &\Lv^{(S+1)} \eth_+^{(S)} - \eth_+^{(S)} \Lv^{(S)} = 0, \label{eq:L_edthp}\\
 &\Lv^{(S-1)} \eth_-^{(S)} - \eth_-^{(S)} \Lv^{(S)} = 0. \label{eq:L_edthm}
\end{align}
\end{subequations}

Using the edth operators \eqref{eq:edth_def}, the Hamiltonian \eqref{eq:Hcal_def} is expressed simply as \cite{Hasebe2015}
\begin{equation}
 \Hcal_\xi = \xi \frac{\hbar v_F}{R} 
 \begin{pmatrix} 0 & -i \eth_{-\xi}^{(\xi S+\frac{\xi}{2})} \\ -i\eth_\xi^{(\xi S-\frac{\xi}{2})} & 0\end{pmatrix}.
\end{equation}
Using Eqs.~\eqref{eq:L_edthp} and \eqref{eq:L_edthm}, one can show 
that the following total angular momentum operator $\Jv_\xi$ commutes with $\Hcal_\xi$: 
\begin{equation}\label{eq:Jv}
 \Jv_\xi = 
 \begin{pmatrix} \Lv^{(\xi S-\frac\xi2)} & 0 \\ 0 & \Lv^{(\xi S+\frac\xi2)} \end{pmatrix}. 
\end{equation}
The unsymmetric form of this operator for the two species arises from 
the effective shifts in the monopole charge due to the spin connection in Eq.~\eqref{eq:pi_def}. 
Using Eqs~\eqref{eq:edth_al_p} and \eqref{eq:edth_al_m}, we find
\begin{equation}
 \Hcal_\xi{}^2=\left(\frac{\hbar v_F}{R}\right)^2 \left( \Jv_\xi{}^2+\frac14-S^2\right), 
\end{equation}
from which we obtain the energy spectrum 
\begin{equation}\label{eq:spec_sph}
 \epsilon_n = \frac{\hbar v_F}{R} \mathrm{sgn}(n) \sqrt{|n|(2S+|n|)},~~ n\in\mathbb{Z}.
\end{equation}
The $n$-th level has the magnitude $j=S-\frac12+|n|$ of the angular momentum $\Jv_\xi$, 
and is $(2j+1)$-fold degenerate. 
The sphere spectrum \eqref{eq:spec_sph} coincides with the disc spectrum \eqref{eq:spec_disc} when $|n|\ll S$. 

The single-particle eigenstates in the ZLL ($n=0$) have the total angular momentum $j=S-\frac12\equiv \St$, and are given by 
\begin{subequations}\label{eq:wf_ZLL_pm}
\begin{align}
&\psi_m^{(+)}(\rv) = \frac{\vb^{\St+m} (-\ub)^{\St-m}}{\sqrt{4\pi R^2 \Ncal_{\St,-m}}}\begin{pmatrix} 1
\\  0\end{pmatrix},\\
&\psi_m^{(-)}(\rv) = \frac{u^{\St+m} v^{\St-m}}{\sqrt{4\pi R^2 \Ncal_{\St m}}}\begin{pmatrix} 1
\\  0\end{pmatrix},
\end{align}
\end{subequations}
for the valleys $\xi = \pm 1$, respectively. 
Here, $m = -\tilde{S}, ..., \tilde{S}$ is the $z$ component of total angular momentum, 
and $\rv$ is constrained to the surface of the sphere ($\rv=R\ev_r$).
We have also introduced the spinor coordinates
\begin{equation}\label{eq:uv}
\begin{split}
 &u= \cos (\theta/2) e^{i\phi/2}\equiv c e^{i\phi/2},\\
 &v= \sin (\theta/2) e^{-i\phi/2}\equiv s e^{-i\phi/2} ,
\end{split}
\end{equation}
and their complex conjugate counterparts $\ub$ and $\vb$, 
where $c$ and $s$ are shorthand notations. 
The normalization factor $\Ncal_{\St m}$ is calculated to be
\begin{equation}\label{eq:Ncal_m}
\begin{split}
\Ncal_{\St m}
&=\int \frac{d^2\rv}{4\pi R^2} c^{2(\tilde{S}+m)} s^{2(\tilde{S}-m)} \\
&=\int_0^\pi d\theta ~c^{2(\tilde{S}+m)+1} s^{2(\St-m)+1}\\
&=\frac{(\tilde{S}+m)!(\tilde{S}-m)!}{(2\tilde{S}+1)!}.
\end{split}
\end{equation}
Both the states in Eq.~\eqref{eq:wf_ZLL_pm} are pseudospin-polarized, and 
the wave functions in the pseudospin-up component are the same as the eigenstates of the non-relativistic lowest Landau level 
with the monopole charge $\xi (2\St)$ \cite{PhysRevLett.51.605,PhysRevB.34.2670,PhysRevA.90.033602}, 
except that $R$ in the normalization of Eq.~\eqref{eq:wf_ZLL_pm} is given by $l_B \sqrt{S}$ rather than $l_B \sqrt{\St}$. 
We note that the states \eqref{eq:wf_ZLL_pm} have average locations 
\begin{equation}\label{eq:cos_th_av}
 \bra{\psi_m^{(\xi)}} \cos\theta \ket{\psi_m^{(\xi)}} 
 =\int \!\! d^2\rv \cos\theta~\psi_m^{(\xi)\dagger}(\rv) \psi_m^{(\xi)}(\rv)=\frac{-\xi m}{\St+1}. 
\end{equation}
In particular, the $m=\St$ state is localized around the south (north) pole of the sphere for $\xi=+~(-)$. 
Such reversed locations between the two valley arise from the fact that the mutually antiparallel pseudomagnetic fields are induced around the two valleys. 

\subsection{Pseudopotentials} \label{sec:pseudopotential}

\newcommand{\Mcal}{{\cal M}}
\newcommand{\Vt}{\tilde{V}}

We consider the interaction Hamiltonian \eqref{eq:H_inter_eff} within the restricted manifold spanned by the ZLL states \eqref{eq:wf_ZLL_pm}. 
In this restricted manifold, the interactions can be conveniently represented in terms of Haldane's pseudopotentials \cite{PhysRevLett.51.605,PhysRevB.34.2670}. 
We calculate the pseudopotentials for both the intra- and inter-valley interactions. 
Expressions of the pseudopotentials for a general interaction potential $V(r)$ are derived in Appendix~\ref{sec:calc_pseudopotential}. 
Here we calculate them for a power-law-decaying potential $V(r)=Cr^{-\alpha}$, in particular, for a dipole-dipole interaction with $\alpha=3$. 
We note that the calculation of the pseudopotentials goes basically in the same way as the non-relativistic case \cite{PhysRevLett.51.605, PhysRevB.34.2670} 
since the ZLL states \eqref{eq:wf_ZLL_pm} have essentially the same form as the lowest-Landau-level states in the non-relativistic case. 

To introduce the pseudopotentials, we first note that because of the spherical symmetry, 
two-body eigenstates for a general interaction potential $V(|\rv_1-\rv_2|)$ are constructed through the angular momentum coupling of Eq.~\eqref{eq:wf_ZLL_pm} as
\begin{equation}\label{eq:Phi_IM}
\begin{split}
 \Phi_{I M}^{(\xi,\eta)}(\rv_1,\rv_2)
 = &\sum_{m_1+m_2=M} \psi_{m_1}^{(\xi)}(\rv_1)\otimes \psi_{m_2}^{(\eta)}(\rv_2)\\ 
 &~~~~~~~~~~\times \bracket{\St,m_1;\St,m_2}{I,M}, 
\end{split}
\end{equation}
where $\bracket{\St,m_1;\St,m_2}{I,M}$ is the Clebsch-Gordan coefficient and $\xi,\eta=\pm$. 
Here, $I$ and $M$ are the magnitude and $z$-component, respectively, of the total angular momentum of the two particles. 
The pseudopotential $V_I^{(\xi,\eta)}$ is defined as the eigenvalue of the interaction potential $V(|\rv_1-\rv_2|)$ for the state \eqref{eq:Phi_IM}. 

Since the interaction Hamiltonian \eqref{eq:H_inter_eff} consists only of two-body scattering processes, 
we can decompose $H_\inter^{(\xi,\eta)}$ in terms of the two-body eigenstates \eqref{eq:Phi_IM} as
\begin{align}\label{eq:H_VPP}
 H_\inter^{(\xi,\eta)} = \frac12 \sum_{I=0}^{2\St} V_I^{(\xi,\eta)} \sum_{M=-I}^I P_{I M}^{(\xi,\eta)\dagger} P_{I M}^{(\xi,\eta)}.
\end{align}
Here, we have introduced the pair creation operator 
\begin{equation}
 P_{I M}^{(\xi,\eta)\dagger} = \sum_{m_1+m_2=M} c_{m_1}^{(\xi) \dagger} c_{m_2}^{(\eta)\dagger} \bracket{\St,m_1;\St,m_2}{I,M}, 
\end{equation}
where $c_m^{(\xi)\dagger}$ is the fermionic creation operator for the $m$-th state \eqref{eq:wf_ZLL_pm} in the ZLL at the valley $\Kv_\xi$. 
For intra-valley interactions $(\xi,\eta)=(+,+)$ and $(-,-)$, 
the sum  in Eq.~\eqref{eq:H_VPP} is restricted to odd $2\St-I$ because of the Fermi statistics. 
Remarkably, while the interaction Hamiltonian  \eqref{eq:H_inter_eff} is originally specified by the continuous function $V(r)$, 
it is now represented by a finite number of parameters, $\{ V_I^{(\xi,\eta)} \}$. 

As described in Appendix~\ref{sec:calc_pseudopotential}, for a general interaction $V(r)$, the pseudopotentials are obtained as
\begin{align}\label{eq:V_I}
 V_I^{(\xi,\eta)}=\frac{[(2\St+1)!]^2}{(2\St-I)!(2\St+I+1)!} \sum_{k=0}^I \frac{(I+k)!}{(k!)^2 (I-k)!} \Vt_k^{(\xi,\eta)},
\end{align}
where 
\begin{subequations}\label{eq:Vk}
\begin{align}
 \tilde{V}_k^{(+,+)} = \tilde{V}_k^{(-,-)} = \int_0^\pi \!\! d\theta ~c^{2k+1} s^{4\St-2k+1}  V(2Rs), \label{eq:Vk_pp}\\
 \tilde{V}_k^{(+,-)} = \tilde{V}_k^{(-,+)} = \int_0^\pi \!\! d\theta ~s^{2k+1} c^{4\St-2k+1}  V(2Rs), \label{eq:Vk_pm}
\end{align}
\end{subequations}
and $c\equiv\cos(\theta/2)$ and $s\equiv\sin(\theta/2)$ as defined in Eq.~\eqref{eq:uv}. 
Below we calculate Eq.~\eqref{eq:V_I} for a power-law-decaying potential $V(r)=Cr^{-\alpha}$. 
We note that integrals in Eq.~\eqref{eq:Vk} diverge for some $k$ owing to a short-distance singularity of $V(r)$, and need a careful treatment. 

We first consider the intra-valley interactions. 
For $4\St-2k+1-\alpha>-1$, i.e., $k<2\St+1-\frac{\alpha}2$, the integral in Eq.~\eqref{eq:Vk_pp} converges, and is calculated to be
\begin{equation}\label{eq:Vk_pp_a}
 \Vt_k^{(+,+)} = \frac{C}{(2R)^\alpha} \frac{k! (2\St-k-\frac{\alpha}2)!}{(2\St+1-\frac{\alpha}2)!}. 
\end{equation}
The integral in Eq.~\eqref{eq:Vk_pp} diverges otherwise. 
Here, the factorial $x!$ for a real number $x>0$ is defined via the Gamma function as $x!=\Gamma(x+1)$. 
For $I<2\St+1-\frac{\alpha}2$, the sum in Eq.~\eqref{eq:V_I} involves only convergent numbers, 
and is calculated to be (see Appendix~\ref{sec:calc_pseudopotential}) 
\begin{equation}\label{eq:V_I_a}
\begin{split}
 &V_I^{(+,+)} = \frac{C}{(2R)^\alpha}\\
 &\times\frac{ [(2\St+1)!]^2 (2\St-I-\frac{\alpha}2)! (2\St+I+1-\frac{\alpha}2)!}{ [(2\St+1-\frac{\alpha}2)!]^2 (2\St-I)! (2\St+I+1)!}.
\end{split}
\end{equation}
For a dipole-dipole interaction ($\alpha=3$), $V_I^{(+,+)}$ diverges for $I=2\St$, 
but this does not correspond to an allowed scattering channel for fermions. 
Thus we can use Eq.~\eqref{eq:V_I_a} for all the allowed scattering channels for the intra-valley interactions. 
We note that for a Coulomb interaction ($\alpha=1$), Eq.~\eqref{eq:V_I_a} coincides with the result in the non-relativistic case in Ref.~\cite{PhysRevB.34.2670}. 

We next consider the inter-valley interactions. 
For $2k+1-\alpha>-1$, i.e., $k>\frac{\alpha}2-1$, Eq.~\eqref{eq:Vk_pm} is calculated to be
\begin{equation}\label{eq:Vk_pm_a}
 \Vt_k^{(+,-)} = \frac{C}{(2R)^\alpha} \frac{(k-\frac{\alpha}2)!(2\St-k)!}{(2\St+1-\frac{\alpha}2)!} .
\end{equation}
For $k\le \frac{\alpha}2-1$, the integral in Eq.~\eqref{eq:Vk_pm} diverges, and we need to regularize it appropriately. 
A natural short-distance cutoff for the interaction potential $V(r)$ is given by the length $a$ of a nearest-neighbor bond introduced in Eq.~\eqref{eq:deltav}. 
Setting $V(R\theta)=0$ for $0\le \theta \le a/R$, we find 
\begin{equation}\label{eq:Vk_pm_cutoff}
\begin{split}
 \Vt_k^{(+,-)} 
 &\simeq \frac{C}{(2R)^\alpha} \int_{a/R} d\theta \left( \frac{\theta}{2} \right)^{-\alpha+2k+1}\\
 &\simeq \frac{C}{(2R)^\alpha} \frac{2}{\alpha-2k-2} \left(\frac{2R}{a}\right)^{\alpha-2k-2}.
\end{split}
\end{equation}
For a dipole-dipole interaction ($\alpha=3$), this cutoff-dependence occurs for $k=0$, 
which contributes to Eq.~\eqref{eq:V_I} for all $0\le I\le 2\St$. 
For the experimental condition considered in Sec.~\ref{sec:exp_real}, we have $a/l_B\approx 0.25$. 
Here we use this ratio in evaluating Eq.~\eqref{eq:Vk_pm_cutoff}. 


\begin{figure}
   \centering
\includegraphics[width = 0.5\textwidth]{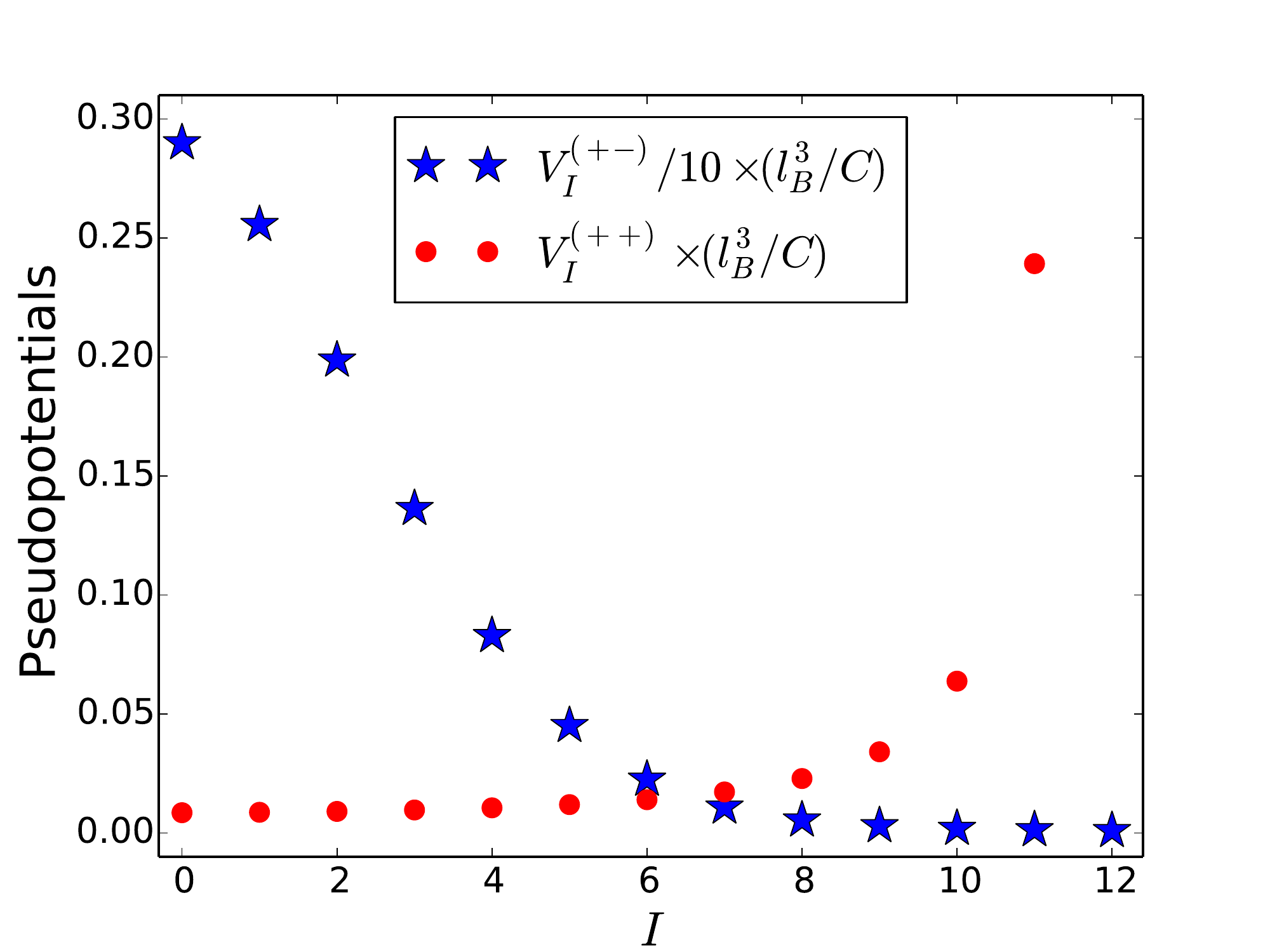}
\caption{(Color online) 
Pseudopotentials \eqref{eq:V_I} for a dipole-dipole interaction $V(r)=C/r^3$ on a sphere with $2\St=2S-1=12$. 
Here, the intra-valley pseudopotentials (red circles) are calculated from Eq.~\eqref{eq:V_I_a}, 
and the inter-valley ones (blue stars; scaled by $1/10$) are obtained by inserting Eqs.~\eqref{eq:Vk_pm_a} and \eqref{eq:Vk_pm_cutoff} into Eq.~\eqref{eq:V_I}. 
For the intra-valley interactions, only the channels with odd (even) $I$ are allowed for even (odd) $2\St$ because of the Fermi statistics. 
The intra-valley pseudopotential with $I=2\St=12$ is not shown because it diverges, 
but in any case does not correspond to an allowed channel for fermions. 
}
          \label{pseudo}
\end{figure}

Setting $2\St=2S-1=12$, we plot the pseudopotentials for the intra- and inter-valley interactions in Fig.~\ref{pseudo}. 
We first find that the inter-valley pseudopotentials have a far larger scale than the intra-valley ones.
This comes from the diverging contribution of the short-distance part of $V(r)$ as found in Eq.~\eqref{eq:Vk_pm_cutoff}, 
and causes a spontaneous valley-polarization as we discuss later in Sec.~\ref{sec:valley_polar}. 
We further find that the intra- and inter-valley pseudopotentials depend differently on $I$: 
the former (latter) monotonically increases (decreases) with increasing $I$. 
This can be understood as follows. 
Equation~\eqref{eq:cos_th_av} indicates that a particle having an average angular momentum $\langle \Jv_\xi \rangle$ 
is localized in the direction of $-\xi \langle \Jv_\xi \rangle$ on the sphere. 
Thus, an intra-valley (inter-valley) repulsive interaction on the sphere implies an ``antiferromagnetic'' (``ferromagnetic'') interaction 
between the angular momenta of the two particles, resulting in a larger energy cost for larger (smaller) $I$. 

\section{Numerical investigation of fractional quantum Hall states}\label{sec. 4}

   \begin{figure}[b]
   \centering
    \includegraphics[width=0.5\textwidth]{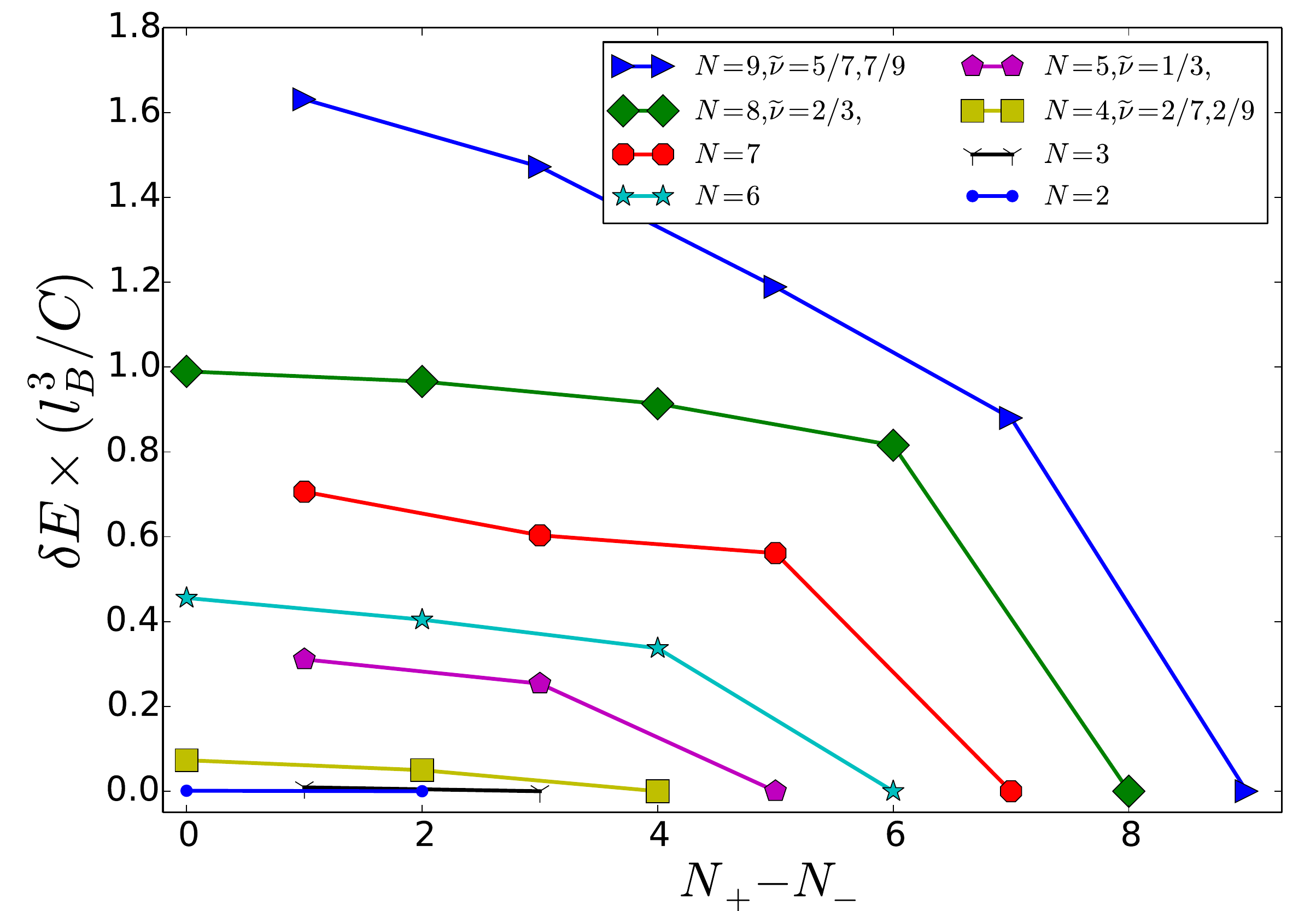}
\caption{(Color online) 
Dependence of the ground-state energy on the population imbalance $N_+-N_-$ between the two valleys 
for a dipole-dipole interaction on a sphere with $2\St=2S-1=12$. 
We have used the pseudopotentials shown in Fig.~\ref{pseudo}. 
For each value of $N=N_+ + N_-$, we determine the ground-state energy $E(N_+,N_-)$ for each distribution $(N_+,N_-)$ of particles at the two valleys, 
and plot $\delta E = E(N_{+},N_{-}) - E(N, 0)$ in units of $C/l_B^3$. 
}
          \label{occupation}
    \end{figure}

We consider the situation in which the ZLL is partially populated as in Fig.~\ref{Landau_levels}, 
and numerically investigate the FQH states stabilized by a dipole-dipole interaction. 
We have performed ED calculations for the interaction Hamiltonian \eqref{eq:H_inter_eff} on a spherical geometry, 
using the pseudopotential representation as described in Sec~\ref{sec:pseudopotential}. 
We demonstrate that owing to the strong inter-valley pseudopotentials, the ground state is spontaneously fully valley-polarized for an arbitrary filling factor. 
We then carry out an extensive search for incompressible states in the valley-polarized case, 
and find that a variety of FQH states such as Laughlin and composite-fermion states are stabilized. 
We estimate the energy gaps above these states, and discuss the possibility of realizing these states in experiments. 

When specifying the filling factor in this section, we use $\nut$ rather than $\nu$ in Eq.~\eqref{eq:nu_N} 
since the former corresponds directly to the filling factor in the non-relativistic case. 
As mentioned at the end of Sec.~\ref{sec. 2}, we focus on the case of $0<\nut<1$ (i.e., $-1<\nu<0$). 

We note that Ref.~\cite{PhysRevLett.108.266801} has analyzed strongly correlated phases in strained graphene 
through ED of a lattice model in a torus geometry. 
Compared to their work, our approach based on a continuum theory on a sphere can significantly simplify 
the search for candidate incompressible states through the use of the total angular momentum of the ground state 
as we demonstrate in Sec.~\ref{sec:search_FQH}. 
Furthermore, there is no topological degeneracy of the ground state on a sphere, 
which simplifies also the analysis of the gap above the ground state. 
Meanwhile, our approach is not suitable in treating short-distance details of interactions on a lattice as in Ref.~\cite{PhysRevLett.108.266801}. 


\subsection{Valley-polarization}\label{sec:valley_polar}

We first demonstrate that the ground state is spontaneously fully valley-polarized for an arbitrary filling factor. 
Figure~\ref{occupation} presents the dependence of the ground-state energy 
on the population imbalance $N_+-N_-$ between the two valleys, for each fixed value of $N=N_+ + N_-$. 
We find that for each $N$, the lowest-energy state is found for the fully imbalanced case $N_+-N_-=N$. 
This indicates that the ground state is spontaneously fully valley-polarized. 
This occurs for all values of the monopole charge $2S$ that we have investigated. 
This can be understood from the enhanced role of inter-valley interactions as found in the behavior of the pseudopotentials in Fig.~\ref{pseudo}. 
We note that a similar behavior has been discussed as a phase separation instability in Ref.~\cite{PhysRevB.85.195113}.


\begin{center}
\begin{figure}[b]
   \includegraphics[width=0.5\textwidth]{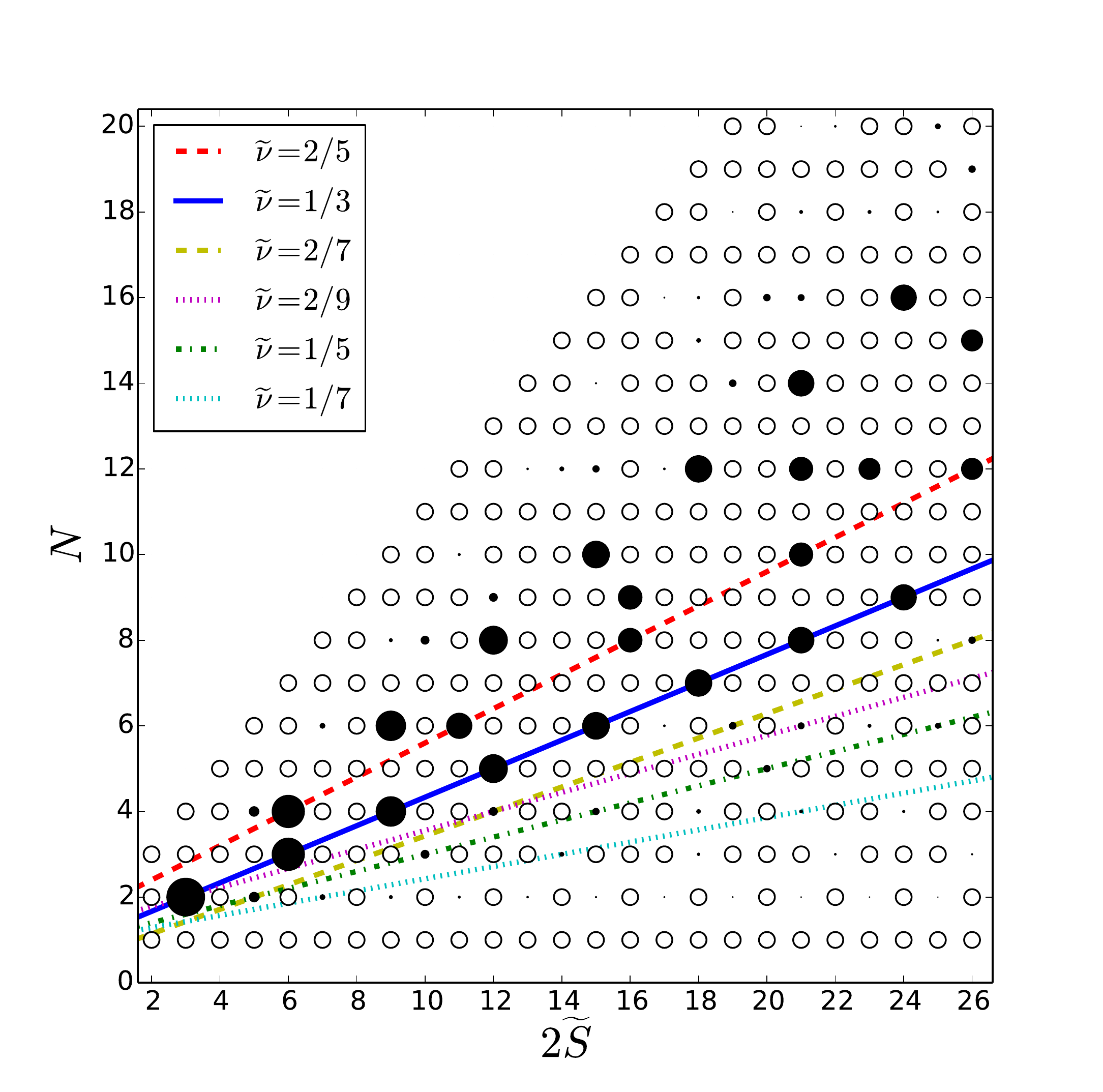}
\caption{(Color online) 
Candidates of incompressible ground states in the $(2\St,N)$ plane for a dipole-dipole interaction. 
The filled circles indicate ground states with the total angular momentum $J=0$, where incompressible states can appear. 
The area of each filled circle is proportional to the neutral gap. 
Lines indicate the relation \eqref{eq:shift} for candidate FQH states listed in the Table \ref{seriesFQH}. 
The empty circles indicates ground states with $J>0$, which are in general compressible. 
%
}
      \label{GS_profile}
\end{figure}
\end{center}

\subsection{Numerical search for FQH states}\label{sec:search_FQH}

Focusing on the fully valley-polarized sector with $(N_+,N_-)=(N,0)$, 
we have carried out an extensive search for incompressible ground states in the $(2\St,N)$ plane as shown in Fig.~\ref{GS_profile}. 
We note that ED in this sector is insensitive to the choice of the short-distance cutoff 
since the intra-valley pseudopotentials $V_I^{(+,+)}$ for $I\le 2\St-1$ do not depend on the cutoff as seen in Eq.~\eqref{eq:V_I_a} 
($V_{2\St}^{(+,+)}$ does depend on the cutoff, but does not correspond to an allowed scattering channel for fermions). 
Furthermore, in this sector, the results are symmetric around $\nut=1/2$ 
because the particle-hole transformation relates the filling factors $\nut$ and $1-\nut$. 

Because of the spherical symmetry, the total angular momentum, 
which is defined as the sum of the $\Jv_\xi$ operator in Eq.~\eqref{eq:Jv} over all the particles, commutes with the Hamiltonian. 
ED calculation can thus be performed separately for different values of the magnitude $J$ of the total angular momentum. 
Incompressible states in general appear as the unique ground states with $J=0$, which are indicated by filled circles in Fig.~\ref{GS_profile}. 
The area of each filled circle is proportional to the neutral gap, which is defined as the excitation gap for fixed $(2\St, N_+,N_-)$. 

In the thermodynamic limit, the filling factor is given by $\nut=N/(2S)=N/(2\St+1)$. 
However, for incompressible states on a finite sphere, 
the relation between $N$ and $2\St$ involves a characteristic shift $\delta$ \cite{PhysRevLett.54.237}: 
\begin{equation}\label{eq:shift}
 N=\nut ( 2\St+\delta ),
\end{equation}
where $\delta$ depends on an individual candidate wave function. 
For example, the Laughlin state at $\nut=1/(2p+1)$ has the shift $\delta=2p+1$. 
The relation~\eqref{eq:shift} can be used to identify different FQH states. 

\begin{table}
\caption{\label{seriesFQH}
FQH states with $0<\nut<1/2$ that can be identified in Fig.~\ref{GS_profile}. 
Each state has the characteristic filling factor $\nut$ and shift $\delta$.  
For each state in the table, at least three filled circles are found to be on the line with \eqref{eq:shift} in Fig.~\ref{GS_profile}. 
All of these states can be interpreted as integer quantum Hall states of composite fermions with $(\nut,\delta)$ in Eq.~\eqref{eq:comp_fer}, 
and the corresponding $p$ and $n$ values are also shown. 
}
\begin{tabular}{ccccc}
\hline\hline 
~~~~$\nut$~~~~ & ~~~~$\delta$~~~~ & ~~~~$p$~~~~ & ~~~~$n$~~ & description \\ 
\hline
  $2/5$ & $4$ & $1$ & ~~$2$ & principal sequence \\
  $1/3$ & $3$ & $1$ & ~~$1$ & Laughlin \\
  $2/7$ & $2$ & $2$ & $-2$ & \\
  $2/9$ & $6$ & $2$ & ~~$2$ & \\
  $1/5$ & $5$ & $2$ & ~~$1$ & Laughlin \\
  $1/7$ & $7$ & $3$ & ~~$1$ & Laughlin \\
\hline\hline
\end{tabular}
\end{table}


FQH states with $0<\nut<1/2$ that can be identified in Fig.~\ref{GS_profile} are summarized in Table \ref{seriesFQH}. 
All the states can be interpreted as integer quantum Hall states of composite fermions \cite{Jain_CompositeFermion, PhysRevLett.63.199, PhysRevB.41.7653}, 
which have the following filling factor and shift: 
\begin{equation}\label{eq:comp_fer}
 \nut=\frac{n}{2pn+1},~~\delta=2p+n. 
\end{equation}
These states include the Laughlin states ($n=1$) \cite{PhysRevLett.50.1395} and Jain's principal sequence ($p=1$) as the special cases. 
We note that the counterparts of the states in Table \ref{seriesFQH} under the particle-hole transformation, 
which have the filling factors $\nut=3/5$, $2/3$, etc., are also seen in Fig.~\ref{GS_profile}. 
We note that the appearance of the Laughlin state at the $1/3$ filling for dipolar fermions 
has also been discussed in Refs.~\cite{PhysRevLett.94.070404,PhysRevA.84.043605,PhysRevLett.99.160403}. 

\begin{figure}
\centering
\includegraphics[width=0.5\textwidth]{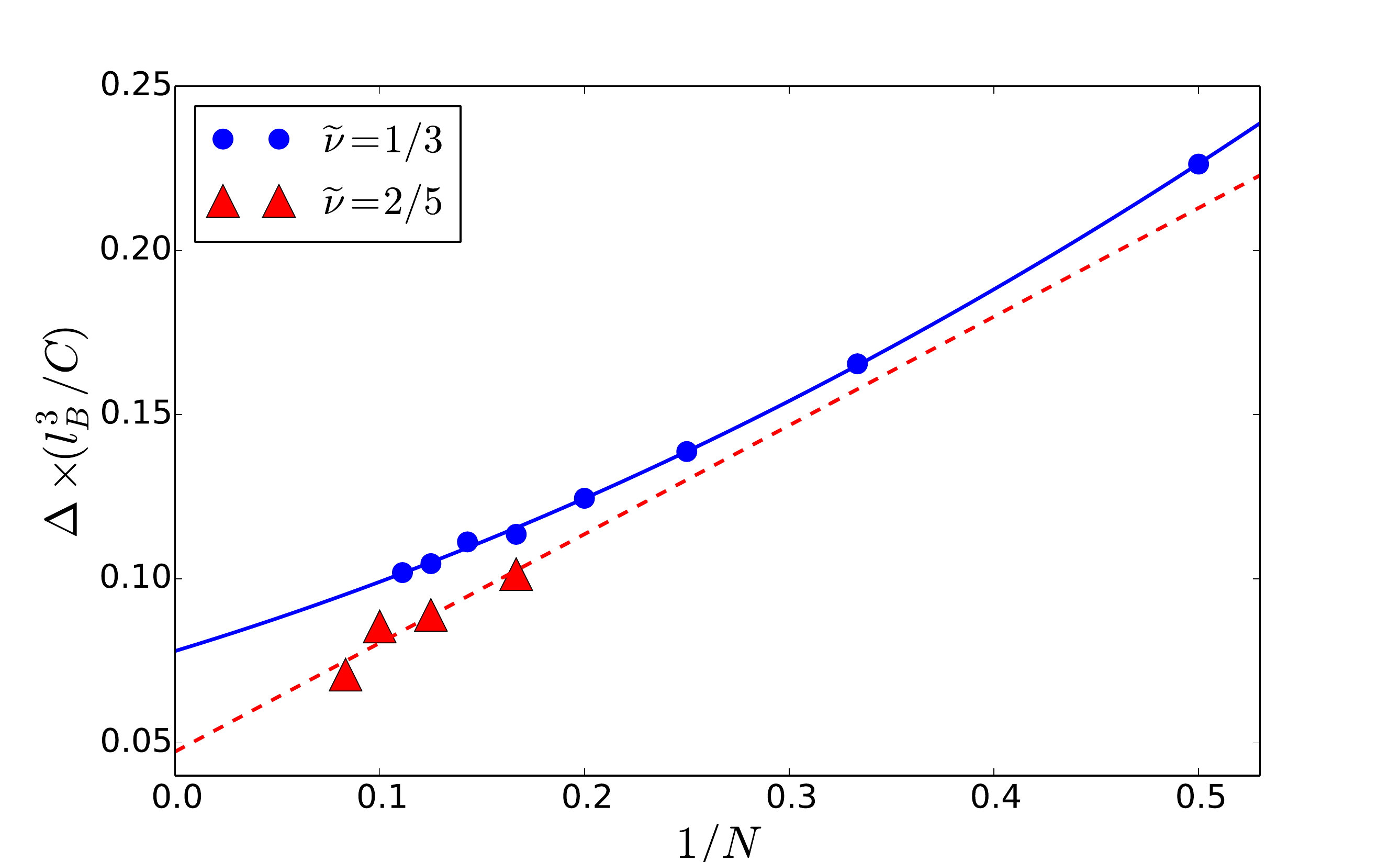}
   \caption{(Color online) Energy gaps $\Delta_{\nut}^{(N)}$ above the ground state as a function of $1/N$ 
   for $\widetilde{\nu} = 1/3$ (blue circles and a solid line) and $\widetilde{\nu} = 2/5$ (red triangles and a dashed line). 
   Symbols and lines indicate the ED data and their fits given by Eq.~\eqref{eq:gap_fit}, respectively. 
}
      \label{fit}
\end{figure}

Among the FQH states listed in Table \ref{seriesFQH}, 
the most prominent gaps are found for the principal sequence $\widetilde{\nu}  =  n/(2n+1)$ with $n = 1, 2$. 
In order to discuss the experimental realizability of these states, we estimate the excitation gaps above the ground states in the thermodynamic limit. 
In Fig.~\ref{fit}, we plot the ED data of the neutral excitation gap as a function of $1/N$. 
For the $\nut=1/3$ state, we fit the data with a quadratic function 
as was done for non-relativistic fermions interacting via a Coulomb interaction \cite{ PhysRevB.34.2670}; 
for the $\nut=2/5$ state, we perform a simple linear fit since the dependence of the data on $N$ is less smooth. 
The fits give the following results: 
\begin{subequations}\label{eq:gap_fit}
\begin{align}
 \Delta_{\nut = 1/3}^{(N)} &= \frac{C}{l_{B}^3}\left(0.0780 + \frac{0.189}{N} + \frac{0.215}{N^2}\right), \label{eq:gap_fit_13}\\
 \Delta_{\nut = 2/5}^{(N)} &= \frac{C}{l_{B}^3}\left(0.0473 + \frac{0.331}{N} \right).
\end{align}
\end{subequations}
In order for $\Delta_{\widetilde{\nu} = 1/3}\equiv \Delta_{\widetilde{\nu} = 1/3}^{(\infty)}$ 
to be the lowest excitation energy above the ground state, 
it must be smaller than the excitation energies to the non-fully-valley-polarized sectors. 
Figure \ref{occupation} indicates that this is true. 
For $N=5$, which corresponds to the $\widetilde{\nu} = 1/3$ FQH state, 
the gap to the sector with $(N_+,N_-)=(N-1,1)$ is given by $\delta E\simeq0.103(C/l_B^3)$; 
this is slightly larger than $\Delta_{\widetilde{\nu} = 1/3}$, 
and should be at least comparable to it even if it is extrapolated to the thermodynamic limit. 

We note that using trial wave functions, 
Ref.~\cite{PhysRevA.84.043605} has estimated the quasihole excitation energy of the Laughlin state at the $1/3$ filling 
to be $ (0.0132\pm 0.0020) C/l_{B}^3$, which is several times smaller than $\Delta_{\widetilde{\nu} = 1/3}$. 
Meanwhile, a natural excitation from the ground state is a ``quasiexciton'' pair of a quasihole and a quasiparticle \cite{PhysRevLett.54.237}. 
The neutral excitation gap estimated in Eq.~\eqref{eq:gap_fit_13} can be interpreted as the lowest excitation energy of such a pair. 
We thus expect that the properties of the Laughlin ground state can be observed if the temperature is below the scale of $\Delta_{\widetilde{\nu} = 1/3}$. 

\begin{figure}
\centering
\includegraphics[width=0.4\textwidth]{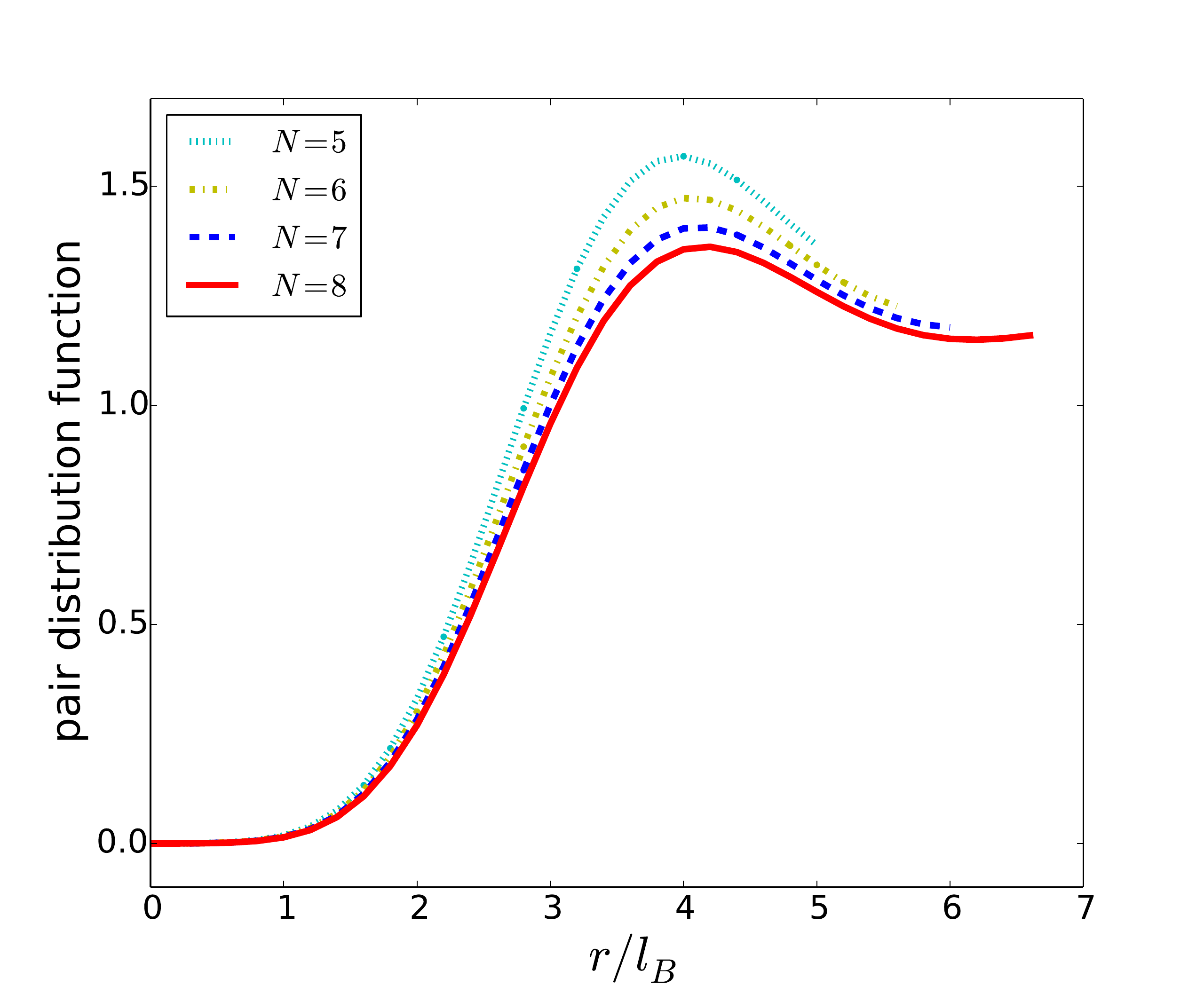}
\caption{(Color online) Pair distribution function $G(r)$ for the ground state at $\nut=1/3$ for different number of particles, $N$. 
}
      \label{fig:PDF}
\end{figure}

For the other FQH states listed in Table \ref{seriesFQH}, 
we do not have a sufficient number of finite-size data to make a reliable extrapolation of the excitation gap to the thermodynamic limit. 
Yet, we find that as we decrease $\nut$, the gap tends to decrease more rapidly than the case of a Coulomb interaction: 
for example, the ratio $\Delta_{\nut=1/5}/\Delta_{\nut=1/3}$ is around $0.05$ and $0.2$ \cite{ PhysRevB.34.2670} for the dipole-dipole and Coulomb cases, respectively 
(we have used the $N=6$ value to estimate $\Delta_{\nut=1/5}$ in the former). 
This can be understood from the behavior of the pseudopotential in Fig.~\ref{pseudo}.   
The intra-valley pseudopotential decreases more rapidly than the Coulomb case as $I$ decreases, 
which reflects the fact that a dipole-dipole interaction decays more rapidly for long distances. 
Specifically, using Eq.~\eqref{eq:V_I_a}, we find $\lim_{\St\to\infty} V_{2\St-3}^{(+,+)}/V_{2\St-1}^{(+,+)}=1/8$ and $5/8$ for $\alpha=3$ and $1$, respectively, 
which is expected to be the main origin of the reduced gap ratio $\Delta_{\nut=1/5}/\Delta_{\nut=1/3}$ in the former case. 
We note that at sufficiently low filling factors $\nut$, the gap would close, leading to the formation of the Wigner crystal \cite{doi:10.1143/JPSJ.72.664}. 
The stability of the Wigner crystal for $\nut<1/7$ has been discussed in Ref.~\cite{PhysRevLett.100.200402}. 
A more detailed analysis of the competition with the Wigner crystal is beyond the scope of the present paper. 


We have also calculated the pair distribution function for the $\nut=1/3$ state; see Fig.~\ref{fig:PDF}. 
For a uniform system of area $\Acal$, it is defined as
\begin{equation}
 G(\rv)=\frac{\Acal}{N(N-1)} \bigg\langle \sum_{i\ne j} \delta (\rv+\rv_i-\rv_j) \bigg\rangle, 
\end{equation}
where $\{\rv_i\}$ are the positions of $N$ fermions and the expectation value is taken with respect to the ground state. 
Because of the spherical symmetry, this function does not depend on the direction of $\rv$ 
and thus $G(\rv)=G(r)$, where $r$ is the chord distance of $\rv$. 
For $\nut=1/3$, a very good approximation to the ground state is given by the Laughlin wave function \cite{PhysRevLett.50.1395, PhysRevLett.51.605}
\begin{equation}
 \Psi_{\nut=1/3} = \prod_{i<j} (\vb_i \ub_j - \ub_i \vb_j)^3. 
\end{equation}
In this wave function, the pair distribution function obeys a power-law dependence $G(r)\propto r^6$ as $r\to 0$. 
This suppression of $G(r)$ for small $r$ marks the effect of a repulsive interaction, 
and can also be found in the numerical data in Fig.~\ref{fig:PDF}. 
As we increase $r$, the numerical data show a hump around $r/l_B=4$, 
and gradually approach unity, which corresponds to the uncorrelated case. 
We thus estimate the correlation length to be around $4l_B$. 



\subsection{Experimental realization}\label{sec:exp_real}

Here we evaluate the energy gap $\Delta_{\widetilde{\nu}=1/3}$ for some experimentally relevant situations. 
We first note that the displacement of the Dirac cones in Eq.~\eqref{eq:f_shift} 
is at most the size of the Brillouin zone: $|\Av(\rv)|/\hbar\lesssim 1/a$. 
Thus the pseudomagnetic field $B/\hbar=1/l_B^2$ obtained from the rotation of $\Av(\rv)/\hbar$ 
is at most of the order of $1/(R_0 a)$ in a sample of radius $R_0$. 
Through a more detailed analysis \cite{PhysRevLett.115.236803}, 
the maximum pseudomagnetic field is estimated to be $B/\hbar=2.7/(R_0 \lambda)$, 
where $\lambda=(3\sqrt{3}/2)a$ is the wavelength of the lasers used to create the honeycomb optical lattice. 
In this case, the gap can be expressed as
\begin{equation}\label{eq:Delta_lamR0}
 \Delta_{\nut=1/3} = 0.0780 \times C \left( \frac{2.7l_B}{R_0 \lambda} \right)^3
 =1.54\times \frac{C}{\lambda^3} \left( \frac{l_B}{R_0} \right)^3
\end{equation}
To achieve a larger gap, it is advantageous to reduce the ratio $R_0/l_B$. 
Meanwhile, the sample radius $R_0$ should be larger than the correlation length $4l_B$ estimated from Fig.~\ref{fig:PDF} 
in order to observe bulk properties around the center of the sample. 

For concreteness, let us consider ${}^{23}$Na${}^{40}$K fermionic polar molecules, 
for which a large electric dipole moment of $d = 0.8$ Debye has been achieved \cite{PhysRevLett.114.205302,ParkWillZwierlein2014}. 
The coefficient of the dipole-dipole interaction is given by $C=d^2/(4\pi\epsilon_0)$, where $\epsilon_0$ is the vacuum permitivity. 
For $\lambda=500$ nm and $R_0/l_B=4$ 
\footnote{In this case, we have $\frac{a}{l_B}=\frac{2.7 l_B}{(3\sqrt{3}/2)R_0}\approx \frac{l_B}{R_0}=0.25$. 
We have used this ratio in the discussion of pseudopotentials in Sec.~\ref{sec:pseudopotential}}, 
the gap is estimated to be $\Delta_{\nut=1/3}\simeq k_B\times 0.89$ nK. 
Since this value is still smaller than the typical temperature scale of ultracold atom experiments, 
we propose to use the recently proposed methods of subwavelength lattices 
\cite{PhysRevLett.111.145304, PhysRevLett.109.235309, PhysRevLett.115.140401,gonzalez2015subwavelength}. 
In this technique, one can create an optical lattice 
whose spacing is reduced by a factor of an integer $N$ \cite{PhysRevLett.115.140401}. 
If we decrease the sample radius $R_0$ by a factor of $N$ at the same time by tightening the trap potential, 
we can keep the ratio $R_0/l_B$ unchanged. 
In this case, Eq.~\eqref{eq:Delta_lamR0} indicates that the gap can be enhanced by a factor of $N^3$. 
If we take $N = 4$, for example, the gap is lifted to about $57$ nK, 
which is in a reasonable range for experimental observation. 
In view of the rapid development in the creation and manipulation of polar molecules, 
molecules with a larger electric dipole moment are likely to be achieved, which would provide another route to a larger gap. 
If FQH states are realized, they can be probed via density plateaus in an {\it in situ} image of the trapped atoms, 
as proposed for integer quantum valley Hall states in Ref.~\cite{PhysRevLett.115.236803}. 


Finally, we comment on the case of magnetic dipolar atoms. 
To be specific, let us consider ${}^{161}$Dy atoms, for which Fermi degeneracy has been achieved \cite{PhysRevLett.108.215301}. 
These atoms have a large magnetic dipole moment of $d=10 \mu_{B}$, where $\mu_B$ is the Bohr magneton. 
The coefficient for the dipole-dipole interaction is given by $C=\mu_0 d^2/(4\pi)$, where $\mu_0$ is the vacuum permeability. 
Comparing this coefficient with that for polar molecules considered above, we find 
$C({}^{161}{\rm Dy})/C({}^{23}{\rm Na}{}^{40}{\rm K})\approx 0.013$. 
Therefore, the excitation gap is two orders of magnitude smaller than the case of polar molecules. 

\section{Summary and outlook}\label{sec. 5}

We have studied strongly correlated ground states of dipolar fermions in a honeycomb optical lattice 
with an effective strain due to spatially varying hopping amplitudes. 
The low-energy effective theory of this system is given by interacting Dirac fermions 
near two valleys in mutually antiparallel magnetic fields. 
We have simulated this theory by ED in the ZLL basis in a spherical geometry. 
In this basis, the interaction Hamiltonian can be conveniently represented in terms of pseudopotentials. 
We have shown that owing to the enhanced inter-valley pseudopotentials, 
the ground state is fully valley-polarized for all the filling factors. 
We have then carried out an extensive search for FQH states in the fully valley-polarized sector, 
and have found signatures of several FQH states which include Laughlin and composite-fermion states of particles and holes. 
The present system can thus emulate FQH physics in a static optical lattice. 
We have calculated the energy gaps above these incompressible states, and discussed the temperature scales required for their experimental realization. 
We have shown that by using the methods of subwavelength optical lattices, we can obtain a reasonable gap for observation. 
We note that the use of Rydberg atoms, 
which have a large dipole moment and strongly interact through enhanced van der Waals force 
\cite{PhysRevA.89.011402,PhysRevLett.104.195302,PhysRevLett.104.223002}, 
may further enhance the energy gap. 


It is interesting to compare the present system with the pseudospin-$\frac12$ Bose gas in antiparallel fields as studied in Ref.~\cite{PhysRevA.90.033602}. 
In Ref.~\cite{PhysRevA.90.033602} (see also Ref.~\cite{PhysRevB.90.245401}),  it was found that fractional quantum spin Hall states 
composed of a pair of nearly independent quantum Hall states are remarkably robust and persist
even when the intercomponent $s$-wave scattering is comparable with the intracomponent one. 
In the present study, we have found that the dipole-dipole interactions of equal magnitudes within each valley and between the valleys 
lead to fully valley-polarized ground states. 
This difference from the bosonic case can be understood from the reduced effect of intra-valley interactions 
due to the prohibition of the scattering channel with $I=2\St$ for fermions. 

In the present work, we have focused on the case of partially filled ZLL as in Fig.~\ref{Landau_levels}. 
By changing the density of the system, one can tune the chemical potential to higher Landau levels with $n=\pm1,\pm 2, ...$, 
and investigate the FQH states realized in those Landau levels. 
It would be interesting to explore the possibility of a non-Abelian quantum Hall state 
as in the case of half-filled second Landau level in GaAs heterostructures \cite{PhysRevLett.59.1776, MOORE1991362, Stern2008204}. 
The realization of a non-Abelian state in a highly controlled setting of ultracold atoms 
would offer a step toward a fault-torelant topological quantum computation \cite{RevModPhys.80.1083}.

\begin{acknowledgments}
HF and YON thank M. Hermanns, E. Ardonne, and S. Nakajima for useful comments.
HF, YON, and SF acknowledge useful discussions during the workshop ``Quantum Information in String Theory and Many-Body Systems'' 
(June, 2016) at Yukawa Institute for Theoretical Physics at Kyoto University. 
HF, YON, and YA were supported by Advanced Leading Graduate Course for Photon Science (ALPS) of 
Japan Society for the Promotion of Science (JSPS). 
The authors were also supported by JSPS KAKENHI Grants No.\ 16J04752, No.\ 16J01135, No.\ 16J03613, and No.\ 23540397. 
\end{acknowledgments}

\appendix

\section{Calculation of pseudopotentials} \label{sec:calc_pseudopotential}
\allowdisplaybreaks

Here we describe some details of the calculation of the pseudopotentials presented in Sec.~\ref{sec:pseudopotential}. 
We essentially follow the method in Ref.~\cite{PhysRevB.34.2670}. 

We fist derive Eq.~\eqref{eq:V_I}, which is an expression of the pseudopotentials for a general interaction potential $V(r)$. 
As described in Sec.~\ref{sec:pseudopotential}, 
the pseudopotential $V_I^{(\xi,\eta)}$ is defined as the eigenvalue of the interaction potential $V(|\rv_1-\rv_2|)$ for the two-body state \eqref{eq:Phi_IM}. 
Since it does not depend on $M$, it is sufficient to calculate it using the highest-weight state with $M=I$: 
\begin{equation}\label{eq:V_I_I}
\begin{split}
 &V_I^{(\xi,\eta)}
 = \bra{\Phi_{II}^{(\xi,\eta)}} V(|\rv_1-\rv_2|) \ket{\Phi_{II}^{(\xi,\eta)}} \\
 &=\int d^2\rv_1 d^2\rv_2 V(|\rv_1-\rv_2|) \Phi_{II}^{(\xi,\eta)\dagger} (\rv_1,\rv_2) \Phi_{II}^{(\xi,\eta)} (\rv_1,\rv_2). 
\end{split}
\end{equation} 
For $M=I$, the following simple expression of the Clebsch-Gordan coefficients is available: 
\begin{equation}\label{eq:CG_II}
\begin{split}
 &\bracket{\St,m;\St,I-m}{I,I} \\
 &=\frac{(-1)^{\St-m}}{I !}  \left[ \frac{(2I+1)! (2\St-I)! (\St+m)! (\St+I-m)!}{(2\St+I+1)! (\St-m)! (\St-I+m)!} \right]^{\frac12}
\end{split}
\end{equation}
Furthermore, because $V_I^{(\xi,\eta)}=V_I^{(-\xi,-\eta)}$, we can focus on the cases of $(\xi,\eta)=(-,-)$ and $(+,-)$. 
Using Eqs.~\eqref{eq:wf_ZLL_pm} and \eqref{eq:CG_II}, the two-body eigenstates \eqref{eq:Phi_IM} are obtained as
\begin{subequations}
\begin{align}
&\Phi_{II}^{(-,-)}(\rv_1,\rv_2)
=\frac{ (u_1v_2-v_1u_2)^{2\St-I}(u_1 u_2)^I }{4\pi R^2\Mcal_{\St I}^{1/2}}
\begin{pmatrix} 1 \\ 0 \end{pmatrix} \otimes \begin{pmatrix} 1 \\ 0 \end{pmatrix} ,\label{eq:Phi_II_mm}\\
&\Phi_{II}^{(+,-)}(\rv_1,\rv_2)
=\frac{ (\vb_1v_2+\ub_1u_2)^{2\St-I} (\vb_1 u_2)^I }{4\pi R^2\Mcal_{\St I}^{1/2}}
\begin{pmatrix} 1 \\ 0 \end{pmatrix} \otimes \begin{pmatrix} 1 \\ 0 \end{pmatrix} ,\label{eq:Phi_II_pm}
\end{align}
\end{subequations}
where the normalization factor $\Mcal_{\St I}$ is given by
\begin{equation}
 \Mcal_{\St I}^{-1/2} = \frac{(2\St+1)!}{I!} \left[ \frac{(2I+1)!}{(2\St-I)! (2\St+I+1)!} \right]^{1/2}, 
\end{equation}
and we have introduced the spinor coordinates $(u_i,v_i)$ for $\rv_i~(i=1,2)$ as in Eq.~\eqref{eq:uv}. 

We calculate the pseudopotentials \eqref{eq:V_I_I} for the intra-valley interaction. 
Substituting Eq.~\eqref{eq:Phi_II_mm} into Eq.~\eqref{eq:V_I_I}, we have
\begin{equation}\label{eq:V_I_mm_uvuu}
\begin{split}
 V_I^{(-,-)}
 &=\int \frac{d^2\Omega_1 d^2\Omega_2}{(4\pi)^2\Mcal_{\St I}} V(2R|u_1v_2-v_1u_2|) \\
 &~~~~~~~~\times|u_1v_2-v_1u_2|^{4\St-2I} |u_1u_2|^{2I},
\end{split}
\end{equation}
where integrations are taken over solid angles $\Omega_i$ formed by $\rv_i~(i=1,2)$. 
We perform the following unitary transformation for the spinor coordinates of the second particle: 
\begin{equation}
 \begin{pmatrix} u_2' \\ v_2' \end{pmatrix} 
 = 
 \begin{pmatrix} \ub_1 & \vb_1 \\ -v_1 & u_1 \end{pmatrix}
 \begin{pmatrix} u_2 \\ v_2 \end{pmatrix}.
\end{equation}
Equation~\eqref{eq:V_I_mm_uvuu} is then rewritten as
\begin{equation}
\begin{split}
 V_I^{(-,-)}
 &=\int \frac{d^2\Omega_1 d^2\Omega_2}{(4\pi)^2\Mcal_{\St I}} V(2R|v_2'|) \\
 &~~~~~~~~\times  |v_2'|^{4\St-2I} |u_1(u_1u_2'-\vb_1v_2')|^{2I}\\
 &=\int \frac{\sin\theta_1\sin\theta_2 d\theta_1 d\phi_1 d\theta_2' d\phi_2'}{(4\pi)^2\Mcal_{\St I}} V(2R s_2') \\
 &~~~~~~~~\times  s_2'^{4\St-2I} c_1^{2I} |c_1c_2'-e^{-i\phi_2'} s_1s_2'|^{2I},
\end{split}
\end{equation}
where $(\theta_2',\phi_2')$ and $(c_2',s_2')$ are defined for $(u_2',v_2')$ as in Eq.~\eqref{eq:uv}. 
The integration over $\phi_1$ trivially yields the constant $2\pi$ while that over $\phi_2'$ gives
\begin{equation}
\begin{split}
&\int_0^{2\pi} \frac{d\phi_2'}{2\pi} |c_1c_2' - e^{-i\phi_2'} s_1s_2' |^{2I}\\
&=\sum_{k=0}^I [C(I,k)]^2 (c_1c_2')^{2k} (s_1s_2')^{2I-2k}, 
\end{split}
\end{equation}
where $C(I,k)$ is the binomial coefficient. 
Therefore, the intra-valley pseudopotentials are calculated as
\begin{equation}\label{eq:V_I_mm_C_Vtk}
\begin{split}
 V_I^{(-,-)} 
 &= \sum_{k=0}^I \frac{[C(I,k)]^2}{\Mcal_{\St I}} \int_0^\pi d\theta_1~ c_1^{2(I+k)+1} s_1^{2(I-k)+1}  \\
 &~~~\times\int_0^\pi d\theta_2'~ c_2'^{2k+1} s_2'^{4\St-2k+1}  V(2Rs_2')\\
 &=\sum_{k=0}^I \frac{[C(I,k)]^2}{\Mcal_{\St I}}  \frac{(I+k)!(I-k)!}{(2I+1)!} \tilde{V}_k^{(+,+)}, 
\end{split}
\end{equation}
which gives Eq.~\eqref{eq:V_I} for $(\xi,\eta)=(-,-)$. 

The inter-valley pseudopotentials can likewise be calculated as
\begin{align}\label{eq:V_I_pm_C_Vtk}
 V_I^{(+,-)}
 &=\int \frac{d^2\Omega_1 d^2\Omega_2}{(4\pi)^2\Mcal_{\St I}} V(2R|u_1v_2-v_1u_2|) \notag\\
 &~~~~~~~~\times |\vb_1 v_2+\ub_1u_2|^{4\St-2I} |\vb_1u_2|^{2I} \notag\\
 &=\int \frac{d^2\Omega_1 d^2\Omega_2'}{(4\pi)^2\Mcal_{\St I}} V(2R|v_2'|) \notag\\
 &~~~~~~~~\times |u_2'|^{4\St-2J} |\vb_1 (u_1u_2'-\vb_1v_2')|^{2I}\notag\\
 &=\int \frac{\sin\theta_1\sin\theta_2 d\theta_1 d\phi_1 d\theta_2' d\phi_2'}{(4\pi)^2 \Mcal_{\St I}} V\left(2R c_2'\right) \notag\\
 &~~~~~~~~\times c_2'^{4\St-2I} s_1^{2I}  |c_1c_2' - e^{-i\phi_2'} s_1s_2' |^{2I}\notag\\
 &= \sum_{k=0}^I \frac{[C(I,k)]^2}{\Mcal_{\St I}}  \int_0^\pi d\theta_1 ~c_1^{2(I-k)+1} s_1^{2(I+k)+1}  \notag\\
 &~~~~~~~~\times\int_0^\pi d\theta_2' ~c_2'^{4\St-2k+1} s_2'^{2k+1}  V(2Rs_2')\notag\\
 &=\sum_{k=0}^{I} \frac{[C(I,k)]^2}{\Mcal_{\St I}} \frac{(I-k)!(I+k)!}{(2I+1)!} \Vt_k^{(+,-)},
\end{align}
which gives Eq.~\eqref{eq:V_I} for $(\xi,\eta)=(+,-)$. 
Equations~\eqref{eq:V_I_mm_C_Vtk} and \eqref{eq:V_I_pm_C_Vtk} complete the derivation of  Eq.~\eqref{eq:V_I}. 

Finally, we note that Eq.~\eqref{eq:V_I_a} is derived using Eq.~\eqref{eq:Vk_pp_a} as follows: 
\begin{align}
V_I^{(-,-)}
&=\frac{I!}{\Mcal_{\St I} (2I+1)!} \sum_{k=0}^I C(I,k) \frac{(I+k)!}{k!} \Vt_k\notag\\
&=\frac{C}{(2R)^\alpha} \frac{(I!)^2 (2\St-I-\frac{\alpha}2)!}{\Mcal_{\St I} (2I+1)! (2\St+1-\frac\alpha2)!} \notag\\
&~~~\times\sum_{k=0}^I C(I,k) \frac{(I+k)!}{I!} \frac{(2\St-k-\frac\alpha2)!}{(2\St-I-\frac\alpha2)!} \notag\\
&=\frac{C}{(2R)^\alpha} \frac{(I!)^2 (2\St-I-\frac{\alpha}2)! (2\St+I+1-\frac\alpha2)!}{\Mcal_{\St I} (2I+1)! [(2\St+1-\frac\alpha2)!]^2} .
\end{align}
Here, the summation has been taken with the following trick:
\begin{align}\allowdisplaybreaks
&\sum_{k=0}^I C(I,k) \frac{(I+k)!}{I!} \frac{(2\St-k-\frac\alpha2)!}{(2\St-I-\frac\alpha2)!}  \notag\\
&= \sum_{k=0}^I C(I,k) \left(\frac{d}{dx}\right)^k (1-x)^{-(I+1)} \notag\\
&~~~~~\times \left(\frac{d}{dx}\right)^{I-k} (1-x)^{-(2\St-I+1-\frac\alpha2)} \bigg|_{x=0}\notag\\
&= \left(\frac{d}{dx}\right)^I (1-x)^{-(2\St+2-\frac\alpha2)} \bigg|_{x=0}\notag\\
&=\frac{(2\St+I+1-\frac\alpha2)!}{(2\St+1-\frac\alpha2)!}.
\end{align}

\bibliography{FQHE}

\end{document}